\documentclass[twocolumn,showpacs,preprintnumbers,amsmath,amssymb]{revtex4}
\usepackage{graphicx}
\usepackage{dcolumn}
\usepackage{bm}
\def\bea {\begin{eqnarray}}
\def\eea {\end{eqnarray}}

\def\be {\begin{equation}}
\def\ee {\end{equation}}

\begin{document}
\title
{Characterizing quark gluon plasma  by thermal photons and lepton pairs 
}
\author{Sukanya Mitra$^1$, Payal Mohanty$^2$, Sabyasachi Ghosh$^1$, 
Sourav Sarkar$^1$ and Jan-e Alam$^1$}
\medskip
\affiliation{$^1$Variable Energy Cyclotron Centre, 1/AF, Bidhan Nagar,
Kolkata - 700064, INDIA
}
\affiliation{$^2$Saha Institute of Nuclear Physics, 1/AF, Bidhan Nagar,
Kolkata - 700064, INDIA
}
\begin{abstract}
The photon spectra measured by the ALICE collaboration in Pb+Pb collisions at Large Hadron Collider (LHC) energies
has been analyzed with a view of extracting the properties of thermal system formed in these collisions. 
The results of the analysis are compared with the previously studied spectra measured at Super Proton Synchrotron (SPS)
and Relativistic Heavy Ion Collider (RHIC) energies.  The thermal dilepton spectra from the Pb+Pb collision
at LHC energy has been predicted for the initial conditions constrained by the thermal photon spectra at the 
same collision conditions.  The slope of the photon has been used to estimate the
rise in the effective degeneracy with the charged particle multiplicity of the system. 
The slopes of the lepton pair spectra for different invariant mass windows have been used to conjecture the
average radial flow velocity of the high temperature phase. 
\end{abstract}

\maketitle
\section{Introduction}
The main aim of colliding heavy nuclei at relativistic 
energies is to create a hot and/or dense thermal phase of 
matter where the quarks and gluons are not
confined inside hadrons 
but move within  nuclear volume. Such a phase of matter
is called quark gluon plasma (QGP). The detection of
QGP in heavy ion collisions (HIC) at RHIC and LHC energies
is one of the most challenging jobs both for experimentalists
and theorists working in this field primarily because of the extremely
transient nature of the QGP. The QGP
evolves dynamically in space and time 
due to high internal pressure,  due to which the system cools 
and reverts to hadronic matter.   
The electromagnetically interacting particles, photons and dileptons
~\cite{mclerran,gale,weldon}(see~\cite{rapp,alam1,alam2} for review) 
are considered to be penetrating probes of the matter 
formed in the collisions of heavy ions at relativistic 
energies because (i) they are produced at each space-time point
of the system and (ii) their mean-free paths are much larger 
than the system-size
and hence bring the information of the production points very efficiently.

The efficiency of photons for being considered as an competent probe
of QGP largely depends on the ability to disentangle the 
photons produced from various stages of evolution of the system 
formed in HIC.  
Therefore, first we  identify the
possible sources of photons above those coming
from the decays of $\pi^0$ and $\eta$ mesons etc, as provided
by the data because 
photons from these decays are already eliminated 
and hence need not be considered in the present analysis. 
Photons produced from the evolving matter under consideration are: 

(i) due to the initial hard collisions of the partons from the nucleons
of the colliding nuclei. 
This contribution may be estimated by using the techniques of perturbative QCD (pQCD)
and the data from pp collisions may be used to validate such calculations. 
The $p_T$ distributions
of photons from proton+proton (pp) collisions at a given
energy can be used as a benchmark for the hard contribution
in HIC. Therefore, estimation of these contributions with minimal
model dependence is important. In view of this,  
in the present analysis
we estimate the hight $p_T$ contributions in HIC by
using the experimental data obtained in pp collisions~\cite{wa98} as
described in section III.

(ii) Photons are also  produced from the interactions of the
{\it `yet-to-be equilibrated'} partons  i.e. from the  
time span between the collision point and the onset of thermalization. 
In case where the thermalization time scale is very small 
the contributions from this interval will be insignificant and 
hence can be neglected.

(iii) Thermal photons originating from the interactions of the
(a) quarks and gluons in the bath and (b) thermal hadrons ($\pi$, $\rho$,
$\eta$, $\omega$, $a_1$ etc).
The estimation of the thermal contribution depends on the
space-time evolution scenario
that one considers.  In case of a deconfinement phase transition, which seems to
be  plausible at RHIC energies (see~\cite{npa757} for a review),
one assumes that QGP is formed initially.
The equilibrated plasma then expands, cools,
and reverts to hadronic matter and finally freezes out
at a temperature, $T_f\sim$, called freeze-out temperature. 
Evidently there will be thermal radiation
from QGP as well as 
from the luminous hadronic fireball which has to be estimated as accurately as
possible in order to have a reliable estimate of the initial temperature.

The momentum distributions of
photons (and dileptons) produced from a thermal system depend on the temperature ($T$)
of the source through the thermal phase space factors of the participants of the
reactions~\cite{cywong}.
Consequently, the transverse momentum ($p_T$) spectra of photon reflects the temperature
of the source.  
For an expanding system 
the situation is, however, far more complex. The thermal phase space factor changes
 - by several factors {\it e.g.} the transverse
kick received by  low $p_T$ photons due to flow originating from the low
temperature hadronic phase (realized when $T<T_c$) populates
the high $p_T$ part of the spectra~\cite{ja1993}.
As a consequence the intermediate or the high $p_T$
part of the spectra  contains contributions from both QGP as well as hadrons.
The photon spectra measured experimentally represents the space-time
integrated yield from the  matter that evolves from an initial hot and dense 
phase to a comparatively cooler and diluted phase of hadronic gas. Therefore, 
the temperature extracted from such spectra will exhibit the average temperature
of the system.  

The experimental data on the $p_T$ distributions of photons from various collision
energies and colliding systems have been analyzed
by using different kinds of models~\cite{ashns,dksbs,huovinen,gallmeister,steffen,peressounko,jarhic,jknbs}
mainly to extract the
temperature of the hot phase formed after the collisions. 
In the present work the slope 
of the photon spectra has been used to estimate the
enhancements of the  effective degeneracy with 
increased multiplicity. The slope of
dilepton spectra for different invariant mass
windows have been used to extract the average radial flow velocity of the high temperature phase. 

In the next section we very briefly describe the various
mechanisms of photon and lepton pair productions.
Section II is dedicated  a compressed description of the space-time dynamics of
the system. The results are presented in section III and section IV is 
devoted to summary and discussions. 
 
\subsection{Sources of thermal photons}
The emission of thermal photons and lepton pairs from QGP and
thermal hadrons have been discussed in earlier works.
In the present work we briefly outline various processes 
for the emissions and refer to the literature for the details.
The rate of thermal photon production per unit space-time volume
per unit four momentum volume
is given by~\cite{mclerran,gale,weldon} (see ~\cite{alam1} for a review):
\begin{equation}
E\frac{dR}{d^3p}=\frac{g^{\mu\nu}}{(2\pi)^3}\,{\mathrm Im}\Pi^R_{\mu\nu}
f_{BE}(E,T)
\label{eq2}
\end{equation}
where ${\mathrm Im}\Pi_{\mu}^{\mu}$ is the
imaginary part of the retarded photon self energy and $f_{BE}(E,T)$ is the
thermal phase space distribution for Bosons. 
For an expanding system, the energy $E$ should replaced by
$u_\mu p^\mu$, where $p^\mu$ and $u^\mu$ are the four
momentum and the hydrodynamic four velocity respectively.

The Hard Thermal Loop~\cite{braaten} approximations has been used by several
authors~\cite{photons} to evaluate the photon spectra originating from  the
interactions of thermal quarks and gluons. 
The complete calculation of emission rate of photons from QGP to order
O$(\alpha\alpha_s)$ has been done
by resuming ladder diagrams in the effective theory~\cite{arnold}, which
has been used in the present work. A set of  hadronic reactions
with all possible isospin combinations have been considered for the 
production of
photons~\cite{npa1,npa2,turbide} from hadronic matter.
The effect of hadronic dipole
form factors has been taken into account in the present
work as in~\cite{turbide}.
\subsection{Sources of thermal dileptons}
The dominant source of the thermal dileptons from QGP 
is the $q\bar{q}$ annihilation ~\cite{cleymans}. 
For the low mass dilepton production from HM the decays
of thermal light vector mesons namely $\rho$, $\omega$ and $\phi$ 
have been considered. The change of spectral function of $\rho$
due to its interaction with $\pi,\omega, a_1, h_1$ (see~\cite{sabya,sabyaEPJC}
for details) and baryons ~\cite{SS_rho_dense}  have been
included in evaluation of lepton pairs from HM. 
For the spectral function $\omega$ the width at non-zero temperature
is taken from Ref.~\cite{SS_om} and no medium effect has
been considered for $\phi$. 
The continuum part of the spectral 
function of $\rho$ and $\omega$ have also been included in the dilepton 
production rate ~\cite{alam2,cont}. 
The model employed in the present work
leads to a good agreement with NA60 dilepton data~\cite{NA60} 
for SPS collision conditions~\cite{SS_JP}.
\section{Expansion Dynamics}
The space time evolution of the system formed in  Pb+Pb 
collisions at $\sqrt{s_{NN}}=2.76$ TeV has been studied by using relativistic hydrodynamics 
with  longitudinal boost invariance~\cite{bjorken} and cylindrical symmetry
~\cite{hvg}.  
We assume that the system reaches the state of 
equilibrium at a time 
$\tau_i$ after the collision. The initial  temperature, 
$T_i$ can be related to the measured hadronic 
multiplicity ($dN/dy$) by the following relation
for system undergoing isentropic expansion:
\begin{equation}
\frac{dN}{dy}= \pi R_A^2 4a_{q}T_i^{3}\tau_i/c
\label{eq4} 
\end{equation}
where $R_A$ is the radius of colliding nuclei, $c$ is a constant $\sim$ 4 and   
 $a_{q}=(\pi^2/90)g_{q}$ where
$g_{q}$ ($=2\times 8+ 7\times 2\times 2\times 3\times N_F/8$) is the 
degeneracy of quarks and gluons in QGP, $N_F$=number of flavours. 
The value of $dN/dy$  
can be calculated from the following equation~\cite{kharzeev}:
\begin{equation}
\frac{dN}{dy}=(1-x)\frac{dn_{pp}}{dy} \frac{<N_{part}>}{2}
+ x\frac{dn_{pp}}{dy}<N_{coll}>
\label{eq5}
\end{equation}
\begin{table}[ht]
\caption{The values of various parameters - thermalization
time ($\tau_i$), initial temperature ($T_i$) 
and hadronic multiplicity $dN/dy$  - used 
in the present calculations.}
\begin{tabular}{lccr}
\tableline
$\sqrt{s_{NN}}$ &2.76 TeV\\ 
centrality  &0-40\%\\
$\frac{dN}{dy}$ &1212\\
$\tau_i$ &0.1 fm\\
$T_i$ &553 MeV\\
$T_c$ &175 MeV\\
$T_f$ &100 MeV\\
EoS &Lattice QCD\\
\tableline
\label{table:1}
\end{tabular}
\end{table}
\begin{table}
\caption{The values the $\langle p_T \rangle$ for different collision 
energies obtained from the direct photon data at low $p_T$ by 
fitting with $a_0\times \exp[-p_T/a_1]$.}
\begin{tabular}{lccr}
\hline
$(1/N_{part})(dN_{ch}/d\eta)$ &$\langle p_T\rangle$\\ 
\hline
1.214 & 245 MeV\\
1.77 & 265 MeV\\
3.47 & 300 MeV\\
\hline
\label{table:2}
\end{tabular}
\end{table}
$N_{coll}$ is the number of 
collisions and contribute to $x$ fraction to the multiplicity $dn_{pp}/dy$ 
measured in $pp$ collision. The number of participants,
$N_{part}$ contributes a
fraction $(1-x)$ of $dn_{pp}/dy$. The values of $N_{part}$ 
and $N_{coll}$  are estimated by using Glauber Model and
the results are in agreement with ~\cite{aamodt}. We have used
$dn_{pp}/dy=4.31$ and $x=0.1$ at $\sqrt{s_{NN}}=$2.76 TeV.
It should be mentioned here that the values of $dN/dy$ (through 
$N_{\mathrm part}$ and $N_{\mathrm coll}$ in Eq.~\ref{eq5})
and hence the $T_i$ (through $dN/dy$ in Eq.~\ref{eq4}) depend 
on the centrality of the collisions. The values of $R_A$ for 
a given centrality has been evaluated by using the relation -  
$R_A\sim 1.1N_{\mathrm part}^{1/3}$.

We use the lattice QCD EoS ~\cite{MILC} for the QGP phase 
and hadronic resonance gas EoS for the hadronic phase~\cite{bmandja}.
The kinetic freeze out temperature, $T_f=100$ MeV is  
constrained by the $p_T$ spectra of hadrons. 
The ratios of various hadrons measured experimentally at
different $\sqrt{s_{\mathrm {NN}}}$
indicate that the system formed in heavy ion collisions chemically decouple
at $T_{\mathrm {ch}} ( > T_f$).  
Therefore, the system remains out of chemical equilibrium
from $T_{\mathrm {ch}}$ to $T_f$. The deviation of the system from
the chemical equilibrium is taken in to account by introducing 
chemical potential for each hadronic species~\cite{hirano}.

\section{Results and Discussion}
\subsection{$p_T$ distributions of photons and dileptons}
\begin{figure}
\begin{center}
\includegraphics[scale=0.4]{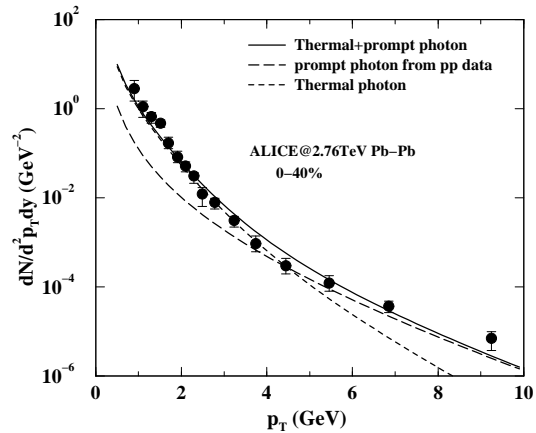}
\caption{Transverse momentum spectra of direct photon  at 2.76 TeV energy for 
Pb+Pb collision  at 0-40\% centrality.}
\label{fig1}
\end{center}
\end{figure} 
\begin{figure}
\begin{center}
\includegraphics[scale=0.35]{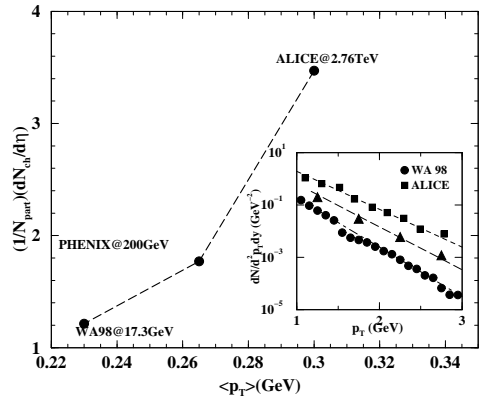}
\caption{The variation of $\langle p_T \rangle$ with 
the increase in multiplicity for different collision energies.}
\label{fig2}
\end{center}
\end{figure} 
\begin{figure}
\begin{center}
\includegraphics[scale=0.3]{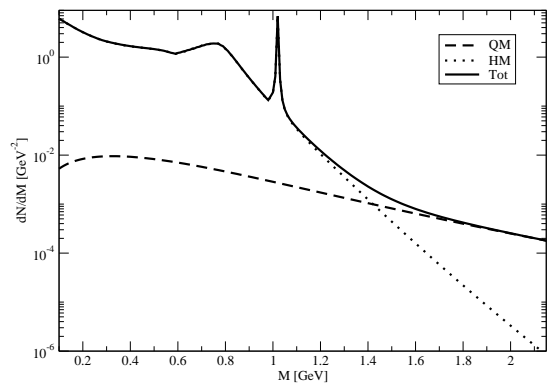}
\caption{The invariant mass distribution of thermal lepton pairs  
for Pb+Pb collision at LHC.} 
\label{fig3}
\end{center}
\end{figure} 
The direct photon spectra from Pb+Pb collisions 
is measured at $\sqrt{s_{NN}}=2.76$ TeV. However,
no data at this collision energy is available for pp interactions. 
Therefore, prompt photons from p+p collision at $\sqrt{s_{NN}}=7$ TeV 
has been used to estimate the hard contributions for
nuclear collisions at $\sqrt{s_{\mathrm NN}}=2.76$ TeV by using 
the scaling (with $\sqrt{s_{NN}}$) procedure used in~\cite{wa98}. 
For the Pb+Pb collisions the result has been scaled up 
by the number of collisions at this energy
(this is shown in Fig.~\ref{fig1} as prompt photons).
The high $p_T$ part of the data is reproduced by the prompt 
contributions reasonably well.  
At low $p_T$ the hard contributions underestimate the data 
indicating the presence of a possible thermal source.

The thermal photons with initial temperature $\sim 553$ MeV 
along with the prompt contributions explain the  data well 
(Fig.~\ref{fig1}),
with the inclusion of non-zero chemical potentials for 
all hadronic species considered~\cite{hirano}(see also~\cite{renk2}). 
\begin{figure}
\begin{center}
\includegraphics[scale=0.3]{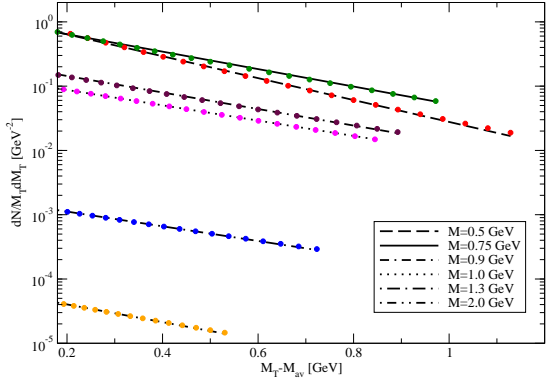}
\caption{The transverse mass distribution of thermal lepton pairs  
for Pb+Pb collision at LHC.} 
\label{fig4}
\end{center}
\end{figure} 
\begin{figure}
\begin{center}
\includegraphics[scale=0.3]{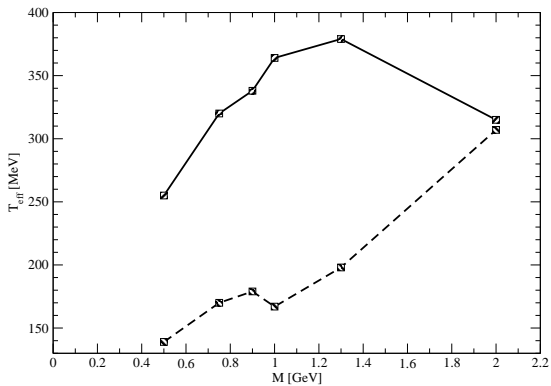}
\caption{The variation of effective slope
with invariant mass bins  
for Pb+Pb collision at LHC. The dashed line
is obtained by setting radial velocity, $v_T=0$} 
\label{fig5}
\end{center}
\end{figure} 

It is well known that transverse momentum spectra of photons act as a
thermometer of the interior of the plasma. The inverse slope of the 
thermal distribution is a measure of the average (over evolution) effective (containing flow) 
temperature  of the system. We have extracted the average effective 
temperature ($\sim <p_T>$) from the thermal distributions of photons at different collision 
energies - {\it i.e.} for SPS, RHIC and LHC energies. Fig~\ref{fig2} shows the 
variation of $\langle p_T \rangle$ with 
multiplicity for different collision energies. 
To minimize the centrality dependence of the results the $dN_{ch}/d\eta$ is 
normalized by $N_{part}$.
The results clearly indicate a significant rise in the  average $p_T$ ($<p_T>$)
while going from SPS to RHIC to LHC. The values of $\langle p_T \rangle$ 
for different collision energies are given in the table~\ref{table:2}.
Since photons are emitted from each space time point of the
system, therefore, the measured slope of the $p_T$ spectra 
represents the average effective  temperature  of the system.

The quantity, $\rho_{eff}^{av}(=1/N_{part}dN_{ch}/d\eta$) is proportional to the
entropy density. Therefore, ${\rho_{eff}^{av}}/{<p_T>^3}\propto g_{eff}^{av}$,
the average effective statistical degeneracy, a
quantity which changes drastically if the colour degrees of freedoms
deconfined {\it i.e.} if a phase transition takes place in the system.
We find that the entropy density ($s\sim g_{eff} T^3$) at LHC increases by almost $96\%$
compared to RHIC and there is an enhancement of $46\%$ at RHIC compared
to SPS. However, part of this increase is due to the increase in the temperature
and part is due to increase in degeneracy. To estimate the increase in the
degeneracy we normalize the quantity $\rho_{eff}^{av}$ by $<p_T>^3$.
Therefore, we estimate $\rho_{eff}^{av}/{<p_T>^3}$ 
from the analysis of the experimental data and 
found that there is a $15\%$ increase in this quantity from SPS to RHIC
and  $35\%$ increase from RHIC to LHC.

We evaluate the invariant mass ($M$) spectra, and the transverse mass ($M_T$)
spectra of lepton pairs with initial conditions, EoS etc constrained 
by the measured photon spectra at $\sqrt{s_{NN}}=2.76$ TeV.  The emission 
processes from QGP and the hadronic phases are taken from Ref.~\cite{sabya},
therefore, we do not repeat the details here to save space.
The $M_T(=\sqrt{M_{av}^2+p_T^2})$ spectra is evaluated for different 
$M$ windows ($M$ ranging from
$M_1$ to $M_2$, with $M_{av}=(M_1+M_2)/2$). The $M$ spectra 
displayed in Fig.~\ref{fig3} indicates that by 
selecting  $M_1$ and $M_2$ appropriately, one
can extract the  properties of QGP ($M_{av}>1.5$ GeV) or hadronic
system ($M_{av}\sim m_\rho$, $\rho$ mass).  Therefore, for example, the slope
of the $M_T$ spectra (Fig.~\ref{fig4}) 
at $M_{av}\sim 2$ GeV and $0.77$ GeV provide information
about the average temperature and flow of the QGP and hadronic phases 
respectively. 

In Fig.~\ref{fig5} the slope of the $M_T$ spectra, $T_{eff}$  for different 
$M$ windows has been plotted. To understand the
effect of flow on the slope of the spectra we evaluate the spectra 
and hence estimate the slope by switching on and off the radial flow. 
The results with (solid line) and without (dashed line) 
flow  are depicted in Fig.~\ref{fig5}. 
The slope, $T_{eff}$ with flow ($v_r\neq 0$) may be
parametrized as $T_{eff}=T_f+M_{av}<v_r>^2$.  The difference
in the slope due to non-zero $v_r$ and the observation of
the dominance of the QGP phase at large $M$ help in
estimating the radial flow of the QGP phase. 
The estimated value of $<v_r>\sim 0.065$ for $M=2$ GeV. 
A very small value of 0.065 is justified because
the $M\sim 2$ GeV range correspond to very early time when 
flow is not fully developed. 
However, for $M\sim 1.6$ GeV we found $<v_r>\sim 0.24$
where QGP phase contributions dominate, indicating the
fact that the QGP formed at LHC collision conditions undergo
significant radial flow.  
Similarly, the value of $<v_r>$ for the hadronic phase
(near the $\rho$ peak) is 0.46. The value of $<v_r>$ 
in the hadronic phase at freeze-out will be more than 0.46 as this is 
the average value of the hadronic matter. 
The $<v_r>\sim 0.37$ for $M\sim 1.3$ GeV  where
the QGP and the hadronic matter contribute almost equally (see Fig.~\ref{fig3}).

\section{Summary and discussions}
In summary, we have analyzed the photon spectra measured by ALICE 
collaboration in Pb+Pb collisions at $\sqrt{s_{NN}}=2.76$ TeV
with a view of extracting the properties of thermal system formed 
in these collisions. 
The slope of the photon spectra have been compared with the previously measured spectra  
at Super Proton Synchrotron (SPS)
and Relativistic Heavy Ion Collider (RHIC) energies.  The thermal lepton 
pair spectra from the Pb+Pb collision
at LHC energy has been estimated with the same initial condition
that is used to reproduce the  
thermal photon spectra at the same collision conditions.  
The increase in the effective statistical degeneracy at RHIC and LHC relative to 
SPS have been estimated from the slope of the photon spectra.
The radial flow velocity of the QGP and hadronic phases have been
assessed from the invariant mass and transverse momentum distributions
of the lepton pairs. As the data from ALICE collaboration is well 
reproduced by the sources of photons described above,
photons from other sources {\it e. g.}  due to jet-thermal parton interactions~\cite{fries}
and induced emissions 
by the hard partons due to multiple interactions in the QGP~\cite{zakharov}
are ignored in the present analysis. These processes do not appear to be 
essential for the present range of transverse momentum. 


\end{document}